\documentclass[twocolumn,showpacs,aps,amssymb]{revtex4}

\usepackage{graphicx}
\usepackage{dcolumn}
\usepackage{bm}

\begin{document}

\title{Gravitational radiation from chiral string cusps}

\author{Eugeny Babichev}
 \email{babichev@inr.npd.ac.ru}
\author{Vyacheslav Dokuchaev}
 \email{dokuchaev@inr.npd.ac.ru}

 \affiliation{Institute for Nuclear Research of the Russian Academy of
Sciences \\
 60th October Anniversary Prospect 7a, 117312 Moscow, Russia}

\date{\today}

\begin{abstract}
We calculate the gravitational radiation from a cusp of the chiral cosmic
strings as a function of the current on the string in the limit of small
values of the current. The smoothing of the cosmic string cusp due to the
presence of the superconducting current on the string leads to the
different behavior of the gravitational radiation from the cusp as
compared with the non-superconducting (ordinary) string. The difference
between the gravitational energy radiation from the chiral cusp and the
ordinary one is proportional to the value of the current on the string.
It is shown that there is a high-frequency cutoff in the spectrum of the
gravitational radiation from the chiral string. The frequency cutoff
depends on the value of the string current. This effect is crucial for
the detection of gravitation wave bursts from cosmic strings: the rather
large current on the string would lower the amplitude of the incoming
signal.
\end{abstract}


\maketitle

\section{\label{sec:intro} Introduction}

Cosmic strings are the linear topological defects that predicted by Grand
Unified Theories (GUT). They may be naturally produced in the phase
transitions in the early universe. A very remarkable feature of the
ordinary (non-superconducting) cosmic string is the presence of the
short-lived cusps, i.~e. the regions of a string moving with the speed
close to the speed of light (see for review e.~g.
\cite{Vilenkin1,Kibble1}). These cusps produce a highly beamed radiation
that can be detected by the gravitational wave detectors
\cite{Damour1,Damour2}.

Witten in 1985 showed that cosmic strings can carry the superconducting
current \cite{Witten1}. The presence of the current on the string leads
to the new interesting features of the string motion: (i) the smoothing of
the cusp and diminishing of their velocities in the case of small
currents; (ii) the possibility for the existence of the stationary cosmic
string configurations (vortons) \cite{Davis1, Haws1, Haws2} in the case of
sufficiently large currents.

The equations of motions of the superconducting cosmic string are
solvable analytically if (i) the current on the string is a chiral one
and (ii) the radiation back reaction is neglected. The motion of the
chiral cosmic string in the appropriate gauge is given as follows
\cite{Carter1, Davis2, Vilenkin3}:
\begin{eqnarray}
\label{x} X^{0}=t, \quad
  \mathbf{X}(t,\sigma)=
  \frac{1}{2}\left[\mathbf{X}_-(\sigma_{-})+\mathbf{X}_+(\sigma_{+})\right],
\end{eqnarray}
where $t$ is the Minkowskian time, $\sigma$ is a parameter along the
string, $\mathbf{X}_-(\sigma_{-})$ and $\mathbf{X}_+(\sigma_{+})$ are
vector functions of $\sigma_{-}=\sigma-t$ and $\sigma_{+}=\sigma+t$
obeying the following constrains:
\begin{equation}
  \label{ab}
  {\mathbf{X}_-'}^{2}(\sigma_{-})=1,\quad
  {\mathbf{X}_+'}^2 (\sigma_{+})=\Delta^2(\sigma_{+})\leq 1,
\end{equation}
where $\Delta(\sigma_{+})$ determines the chiral current along the
string: the more the value of the $\Delta(\sigma_{+})$ the less the
current and vise versa. The smallness of chiral current means that
$\epsilon(\sigma_{+})\equiv 1-\Delta(\sigma_{+})\ll 1$.

Numerical computations of the electromagnetic and gravitational radiation
from chiral cosmic strings as a function of current were done in
\cite{Babichev1}. The limiting case of radiation from cosmic string
strings with a current close to the maximum value was considered in
\cite{Babichev2}. An opposite case for the electromagnetic radiation of
strings with the small chiral currents considered by Blanco-Pillado and
Olum \cite{Blanco}. Damour and Vilenkin \cite{Damour1,Damour2} considered
the gravitational radiation from the cusp of the ordinary strings and
argued that corresponding Gravitational Wave Bursts (GWB) can be detected
by gravitational wave detectors LIGO, VIRGO and LISA. In this paper we
find the gravitational radiation from the cusp of cosmic string in the
limit of the small chiral current. We show that the chiral string cusp
does not radiate the gravitational waves on the frequencies higher than
some cutoff frequency $\omega_{\rm cut}$ which depends on the current
value. Therefore the presence of the current on the string may impede the
observability of the GWBs from the cosmic string cusps by gravitational
wave detectors. We estimate the minimal chiral current leading to the
lowering of amplitude of the gravitational wave on the given frequency
from a single cusp and from the whole string network in the universe. We
calculate also the dependence of full radiated gravitational energy from
the cusp as a function of the small chiral current.

Throughout the paper we adopt the following conventions. We use the units
$\hbar=c=1$. The signature of the space-time is ($-,+,+,+$). Greek indexes
$\mu$, $\nu,\ldots$ run through $0$ to $3$ and Latin labels designate the
three spatial dimensions.

The paper is organized as follows: In Section II we find the dependence of
waveforms for chiral string cusp on the value of the small current on the
string and the angle of deviation of observer from the direction of the
cusp motion. Then we discuss the influence of the presence of the chiral
current on the amplitude of incoming GWBs from the whole universe. In
Section III we make the detailed calculation of the gravitational energy
radiation from the chiral cusp as a function of the small current. In
Section IV we briefly discuss the obtained results.

\section{Waveforms from chiral cusps}

In the following we will consider the case of small current on the string
(i.~e. when $\Delta$ in (\ref{ab}) is close to unity), otherwise cusps do
not formed on the string at all. We also assume that the current is
constant in the part of the string forming the cusp (it means that
$\Delta=const$). Throughout this section we will follow the calculations
of Damour and Vilenkin \cite{Damour1,Damour2}, who considered the
radiation from the ordinary (non-superconducting) string cusp.

Let us denote $X_+^0=\sigma_+$, $X_-^0=-\sigma_-$. Then the solutions of
equations of motion of the chiral string (\ref{x}) can be rewritten in
the following form:
\begin{equation}
\label{X+-}
X^\mu=\frac{1}{2}\left[X_-^\mu(\sigma_-)+X_+^\mu(\sigma_+)\right],
\end{equation}
where $X_-^i(\sigma_-)$ and $X_+^i(\sigma_+)$ are the three-dimensional
components of vector functions $\mathbf{X}_-(\sigma_{-})$ and
$\mathbf{X}_+(\sigma_{+})$ correspondingly. The energy-momentum tensor of
the chiral cosmic string is given by the following expression:
\begin{equation}
\label{T} T^{\mu\nu}=\mu\int\! d t\, d\sigma\,(\dot{X}^\mu
\dot{X}^\nu - X'^\mu X'^\nu)\,\delta^{(4)}(x-X(t,\sigma)),
\end{equation}
where $\mu$ is the energy of the string per unit length.
Inserting (\ref{T}) into general expression for Fourier-transform
of the energy-momentum tensor:
\begin{equation}
\label{Tdef} T^{\mu\nu}(\omega,\mathbf{k})=\frac{1}{2\pi}\int\! d
t \int\! d^3 x\, e^{i(\omega
t-\mathbf{k}\mathbf{x})}T^{\mu\nu}(t,\mathbf{x}) \nonumber
\end{equation}
and using $d t d\sigma=(1/2)d\sigma_+ d\sigma_-$ we obtain the
Fourier-transform of the energy-momentum tensor for chiral cosmic
string:
\begin{equation}
\label{TF}
T^{\mu\nu}(\omega,\mathbf{k})=-\frac{\mu}{4\pi}I^{(\mu}Y^{\nu)},
\end{equation}
where $I^{(\mu}Y^{\nu)}=(1/2)(I^{\mu}Y^{\nu}+I^{\nu}Y^{\mu})$ and
\begin{eqnarray}
  \label{IY}
  I^{\mu}\equiv
  \int d\sigma_- X'^{\mu}_-e^{-\frac{i}{2}k X_-},\,
  Y^{\mu}\equiv
  \int d\sigma_+ X'^{\mu}_+e^{-\frac{i}{2}k X_+},
\end{eqnarray}
where $k=(\omega,\mathbf{k})$. The small perturbations of metric
$h_{\mu\nu}$ for any relativistic source in the linearized
approximation is given by \cite{Weinberg}:
\begin{equation}
\label{h} \bar{h}^{\mu\nu}(t,\mathbf{n})=\frac{4G}{r}\!\int\! d\omega\,
e^{-i\omega(t-r)}\,T^{\mu\nu}(\omega,\mathbf{k}),
\end{equation}
where $r$ is a distance from the source, $\mathbf{n}=\mathbf{x}/r$ and
$\bar{h}^{\mu\nu}\equiv
{h}^{\mu\nu}-(1/2)\eta^{\mu\nu}h^{\lambda}_{\lambda}$. For description of
the gravitational radiation from cusps it is more convenient to use
waveforms
$\kappa^{\mu\nu}(t-r,\mathbf{n})=\bar{h}^{\mu\nu}(t,\mathbf{n})/r$
instead of $h^{\mu\nu}$:
\begin{equation}
\label{kappa} \kappa^{\mu\nu}(t-r,\mathbf{n})=4G\!\int\! d\omega\,
e^{-i\omega(t-r)}\,T^{\mu\nu}(\omega,\mathbf{k}).
\end{equation}
The Fourier-transform of the waveform:
\begin{equation}
\label{kappaF} \kappa^{\mu\nu}(\omega,\mathbf{n})=\!\int \!d
t\,\kappa^{\mu\nu}(t)\,e^{i\omega t}.
\end{equation}
Inserting (\ref{kappa}) into (\ref{kappaF}) and using the
expression for the Fourier-transform of the energy-momentum
tensor (\ref{TF}) we find:
\begin{equation}
\label{kappa1} \kappa^{\mu\nu}(\omega,\mathbf{n})=-2\,G\mu\,
I^{(\mu}Y^{\nu)}.
\end{equation}
For the energy-momentum tensor the following relations are valid:
$k^{\mu}T^{\mu\nu}(\omega,\mathbf{k})=0$ \cite{Weinberg},
therefore using (\ref{kappa1}) and (\ref{TF}) one can see that
the $00$ and $0j$ components of $\kappa^{\mu\nu}$ can be
expressed through the purely spatial components of
$\kappa^{\mu\nu}$. By this reason we will be interested in the
following only in the $\kappa^{ij}$.

Now let us consider the radiation from the relativistic cusp. We
take the following coordinates of the cusp: $\sigma_{\pm}^{0}=0$
and $X_{\mu}^{0}=0$. It is convenient to define the cusp as the
point on the two-dimensional world-sheet of the string where
$\mathbf{X}'_-$ and $\mathbf{X}'_+$ are anti-parallel
\cite{Blanco}:
\begin{equation}
\label{a-par}
\mathbf{n}^{(c)}=-\frac{\mathbf{X}'_-}{|\mathbf{X}'_-|}=
\frac{\mathbf{X}'_+}{|\mathbf{X}'_+|},
\end{equation}
where $\mathbf{X}_-$ and $\mathbf{X}_+$ in (\ref{a-par}) is taken
in at $\sigma_{\pm}=0$ and $\mathbf{n}^{(c)}$ is the direction of
the cusp motion (i.~e. the concentrating of the gravitational
radiation). The position of the string near the cusp can be
expressed as follows:
\begin{eqnarray}
\label{Taylor} \mathbf{X}_-(\sigma_-)&=&-\mathbf{n}^{(c)}\sigma_-
+\frac{1}{2}\mathbf{X}_-''\sigma_-^2 + \frac{1}{6}\mathbf{X}_-'''
\sigma_-^3,\nonumber\\
\mathbf{X}_+(\sigma_+)&=&\Delta\mathbf{n}^{(c)}\sigma_+
+\frac{1}{2}\mathbf{X}_+''\sigma_+^2 + \frac{1}{6}\mathbf{X}_+'''
\sigma_+^3,
\end{eqnarray}
where the derivatives of ${X}^{\mu}_{\pm}$ is taken at
$\sigma_{\pm}=0$. Further, using the conditions (\ref{ab}) one
can find:
\begin{eqnarray}
\label{xl} \mathbf{n}^{(c)} \mathbf{X}''_{\pm}&=&0,\nonumber\\
\mathbf{n}^{(c)} \mathbf{X}'''_{-}&=&(\mathbf{X}''_-)^2,\quad
\mathbf{n}^{(c)}
\mathbf{X}'''_{+}=-\frac{(\mathbf{X}''_+)^2}{\Delta}.
\end{eqnarray}
Damour and Vilenkin \cite{Damour1,Damour2} found for the ordinary
string cusp the qualitative dependence of waveform from the
$\theta$, angle between $\mathbf{n}^{(c)}$ and direction of
emission $\mathbf{n}$. It was obtained that the waveform can be
approximated by the power dependence from the frequency
$\kappa\propto 1/\omega^{4/3}$ for $\omega\alt\omega_{\rm
cut}(\theta)\propto 1/\theta^3$. Note that the dependence of
$\kappa$ on $\omega$ as $\kappa\propto 1/\omega^{4/3}$ differs
from the result in \cite{Damour1,Damour2} because our definition
of $\kappa$ (\ref{kappaF}) does not contain additional $\omega$.
It was found that for large frequencies $\omega\agt \omega_{\rm
cut}$ the waveforms are exponentially suppressed therefore one
can set $\kappa\simeq 0$ for $\omega\agt \omega_{\rm cut}$.

Our aim now is to find the dependence of $\kappa^{ij}$ not only on the
angle $\theta$ but also on the string current. Following the analysis
made in \cite{Damour1,Damour2} we insert (\ref{Taylor}) into (\ref{IY})
and use the conditions (\ref{xl}). We find that function $Y$ defined in
(\ref{IY}) can be rewritten as follows (we omit 4-dimensional indexes
$\mu$ as they are not significant for such an analysis):
\begin{equation}
\label{Y0} Y \propto \int\! d u\,(\varepsilon + u)e^{-i\phi_+},
\end{equation}
where
\begin{equation}
\label{phi0} \phi_+\equiv \frac{1}{2}k X_+\sim u^3+\varepsilon
u^2 + \frac{\theta^2+2\epsilon}{\theta^2}\varepsilon^2 u.
\end{equation}
Here we made the replacement $u=[\omega (X_+'')^2]^{1/3}\sigma_+$
and value $\varepsilon=\theta(\omega/|X_+''|)^{1/3}$ was
introduced. The main difference from the case of the ordinary
string is the presence of the additional multiplier
$(\theta^2+2\varepsilon)/\theta^2$ in (\ref{phi0}) for
$\varepsilon^2 u$. One can see that in the case of sufficiently
small angles $\theta\alt\epsilon$ the behavior of $\phi_+$ is
different from that for ordinary strings. Now the condition
determining the properties of integral (\ref{Y0}) (i.e. when this
integral become exponentially small) is
\begin{equation}
\label{vareps}
\sqrt{\frac{\theta^2+2\varepsilon}{\theta^2}}\,\varepsilon\simeq 1
\end{equation}
instead of $\varepsilon\simeq 1$ for ordinary string cusp
\cite{Damour1,Damour2}. This leads to the different expression for the
cutoff frequency: $\omega_{\rm cut}(\theta,\epsilon)\propto
1/(\theta^2+\epsilon)^{3/2}$, where $\epsilon=1-\Delta\ll 1$. Therefore,
the cutoff frequency $\omega_{\rm cut}$ for small angles ($\theta\alt
\sqrt{\epsilon}$) does not depend on the $\theta$ and proportional to
$\epsilon^{-3/2}$. For other values of $\theta$ we can take that the
radiation from chiral string cusp is equal to that of the ordinary string
cusp, i.~e. $\kappa(\epsilon,\theta)\simeq\kappa(0,\theta)$. Thus we see
that in the case of chiral string cusp the radiation is very small for
frequencies $\omega \agt {\rm \omega_{\rm cut}}$, where
\begin{equation}
\label{cut0} \omega_{\rm cut}= \frac{1}{L\epsilon^{3/2}},
\end{equation}
even when the observer is placed in the direction of the cusp.

Let us now consider the problem of the detection of the gravitational
bursts from chiral string cusps. Damour and Vilenkin argued
\cite{Damour1,Damour2} that the gravitational bursts from ordinary string
cusps can be detected by the gravitational wave detectors. They find that
it is quite possible for LIGO, VIRGO and especially for LISA to detect
the wave bursts from ordinary strings. How does change the inferences
about the detection of GWB in the case of the presence of the chiral
current on the string? Let us suppose for simplicity that all strings in
the universe carry the same small but non-zero current (that is
$\epsilon=const\ll 1$). The main difference of the radiation of chiral
cusp from the radiation of ordinary cusp is the existence of cutoff
frequency (\ref{cut0}), such that signal from cusp with frequencies
$\omega\agt\omega_{\rm cut}$ is highly suppressed. In the simplified
model of string network evolution adopted by Damour and Vilenkin
\cite{Damour1,Damour2} the typical length of the closed strings is given:
\begin{equation}
\label{length} L\sim \Gamma G \mu\, t,
\end{equation}
where the dimensionless coefficient $\Gamma\sim 100$ determines the
gravitational radiation from a string ($\dot{E}=\Gamma G\mu^2$) and $t$
is the cosmological time. It is assumed that signals from cusps reaches
the Earth from different distances that corresponds to different times
$t$. It is more convenient to work with redshifts $z$ rather than with
times $t$. The cosmological time $t$ is connected with the redshift $z$
in a different way in dependence on what epoch we consider: radiation or
matter dominated. Let $z_{\rm eq}$ be the redhift of equal matter and
radiation densities then we have:
\begin{equation}
\label{era} t=\left\{\begin{array}{lcl} t_0 (1+z)^{-3/2},& & z\alt
 z_{\rm eq}; \\
 t_0 (1+z_{\rm eq})^{1/2}(1+z)^{-2},& & z\agt z_{\rm eq}, \\
\end{array} \right.
\end{equation}
where $t_0\simeq 10^{18}$ s is the present age of the universe. To find
the cutoff frequency of incoming signal $\omega_{\rm cut}$ one should
insert successively the time $t$ (\ref{era}) into (\ref{length}), then
(\ref{length}) into the expression for cutoff frequency (\ref{cut0}). And
finally we multiply the resulting expression by the additional factor
$(1+z)$ which takes into account the gravitational signal redshift. As a
result we find:
\begin{equation}
\label{cut} \omega_{\rm cut}=\left\{
\begin{array}{lcl}
(1+z)^{1/2}\left[\Gamma G\mu t_0 \epsilon^{3/2}\right]^{-1}, & &
z\alt z_{\rm eq}; \\
 (1+z)\left[\Gamma G\mu t_0 (1+z_{\rm eq})^{1/2}
 \epsilon^{3/2}\right]^{-1}, & &  z\agt z_{\rm eq}. \\
    \end{array} \right.
\end{equation}
We see that, in any way, matter or radiation dominated era is
considered, the typical length of loops decreases with $z$ and
consequently the cutoff frequency increases with redshift.

The analysis given by Damour and Vilenkin \cite{Damour1,Damour2}
leads to the following result: for a given frequency $\omega$ and
for a given string parameter $\mu$ (or, strictly speaking, for a
given parameter $\Gamma G \mu$) the amplitude of incoming signals
$h$ from cusps in the universe decrease with $z$, but the number
of such signals per unit time $\dot{N}(\mu,\omega)$ increase with
$z$. That is we should find the compromise between the rate of
signals and their amplitude. For a given $\dot{N}$ one can find
the minimal amplitude of incoming signals $h(\mu,\dot{N},\omega)$
and corresponding maximal redshift $z_m(\mu,\dot{N},\omega)$ (in
fact, because $d\dot{N}(z)/dz$ increase with $z$ we can think for
simplicity that all signals come from $z_m$ and have the same
amplitude corresponding to $z_m$ \cite{Damour1,Damour2}).

How does the existence of the cutoff frequency (\ref{cut}) modify these
arguments? The answer depends on the correlation of given frequency
$\omega$ and the cutoff frequency $\omega_{\rm cut}$ (\ref{cut0}) with
$z=z_m(\mu,\dot{N},\omega)$. If $\omega_{\rm cut}\agt\omega$ then the
result does not change, i.e. the amplitude of incoming signal from chiral
string cusp is the same as in the case of ordinary string. In the
opposite case $\omega_{\rm cut}\alt\omega$ the strings with redshift
$z_m(\mu,\dot{N},\omega)$ does not radiate the gravitational waves on
frequency $\omega$. However the dependence of $\omega_{\rm cut}$ on $z$
allows us to shift to a higher $z$ until the relation $\omega_{\rm
cut}=\omega$ is satisfied. This corresponds to a some new (higher)
redshift $z_{\rm cut}$ and consequently to a new (smaller) amplitude of
the incoming signal $h(\mu,\dot{N},\omega_{\rm cut})$. This means that
current may effectively lower the amplitude of GWBs from cosmic string
cusps in the whole universe.

Let us estimate the minimal current on the string that may affect
on the amplitude $h$. There are three regimes of behavior of
$z_m$ in dependence of function $y(\mu,\dot{N},\omega)$:
\begin{equation}
\label{y} y(\mu,\dot{N},\omega)=10^{-2}(\dot{N}/c)t_0^{5/3}
(\Gamma G\mu)^{8/3} \omega^{2/3},
\end{equation}
here $c$ is the the average number of cusps per loop period
(usually it is taken that $c\sim 0.1$). Maximal redshift $z_m(y)$
can be expressed as follows \cite{Damour1,Damour2}:
\begin{equation}
\label{zm} z_m(y)=\left\{\begin{array}{lcl} y^{1/3}, & & y\alt1,\\
 y^{6/11}, & & 1\alt y\alt y_{\rm eq},\\
 (y_{\rm eq}\,y)^{3/11}, & & y_{\rm eq}\alt y,\\
 \end{array}
\right.
\end{equation}
where $y_{\rm eq}=z_{\rm eq}^{11/6}$. Inserting (\ref{zm}) in (\ref{cut})
we can find from the equation $\omega_{\rm cut}=\omega$ the critical
value $\epsilon_{\rm cut}$ when the current become to lower the incoming
amplitude of signal from the cusp. We obtain:
\begin{equation}
\label{ecut} \epsilon_{\rm cut}= \left\{\begin{array}{lcl} \left[
\Gamma G \mu t_0 \omega \right]^{-2/3},& & y\alt 1; \\
\left[10^{-2}(\dot{N}/c)/(\Gamma G \mu t_0^2 \omega^3)
\right]^{2/11},& & y\agt 1.\\
\end{array}
\right.
\end{equation}
Let us fix the rate of observable GWBs $\dot{N}/c\sim 1$, the frequency
$\omega\sim 10^2$ (preferable for detecting of GWB on LIGO/VIRGO
\cite{Damour1,Damour2}) and find $\epsilon_{\rm cut}$ as a function of
$G\mu$. From (\ref{ecut}) we obtain:
\begin{equation}
\label{ecut1} \epsilon_{\rm cut}\sim \left\{\begin{array}{lcl}
10^{-14}(G\mu)^{-2/3},& & G\mu\alt 10^{-10}; \\
10^{-10}(G\mu)^{-2/11},& & G\mu\agt 10^{-10}. \\
\end{array}
\right.
\end{equation}
For example for the string with $G\mu\sim 10^{-12}$ we have $\epsilon_{\rm
cut}\sim 10^{-6}$; if we take $G\mu\sim 10^{-10}$ (corresponding to
$z_m\sim 1$) then $\epsilon_{\rm cut}\sim 10^{-8}$; for usually accepted
$G\mu\sim 10^{-6}$ for GUT string we find $\epsilon_{\rm cut}\sim
10^{-9}$. It may seems that such small values of $\epsilon_{\rm cut}$
corresponds to small currents, but in fact the physical current is
expressed through $\epsilon$ as follows (for $\epsilon\ll 1$):
\begin{equation}
\label{current} j=\sqrt{2 \mu}\sqrt{\epsilon},
\end{equation}
what corresponds to $j_{\rm cut}\sim 10^{10}$ GeV for $G\mu\sim 10^{-10}$
and $j_{\rm cut}\sim 10^{11}$ GeV for $G\mu\sim 10^{-6}$. Note, that
current on the string, generated due to the oscillation of the string
loop in the magnetic field in the universe, is in fact much smaller,
$j_B\sim 10^7$ GeV \cite{Berezinsky}.

\section{Cusp gravitational radiation}

In this section we calculate the full gravitational radiation of
the energy from the single cusp and obtain the dependence of
radiation from the current along the string. The correct behavior
of radiated gravitational energy as function of $\epsilon$ one
can obtain from the following qualitative consideration. For the
radiated gravitational energy we have
\begin{equation}
\label{estE} E(\epsilon)\propto\int\!\theta\, d\theta\!\int\!
\omega^2\, d\omega\, \kappa^2(\epsilon),
\end{equation}
therefore the difference between the radiated energy from the
ordinary and the superconducting string cusps is proportional to:
\begin{equation}
\label{diffE} \delta
E\propto\!\int_0^{\sqrt{\epsilon}}\!\!\theta\,d\theta
\!\int_0^{\infty}\!\! \omega^2\,
d\omega\left[\kappa^2(0)-\kappa^2(\epsilon)\right]\propto
\sqrt{\epsilon}.
\end{equation}
(In the last estimation the result for cutoff frequency
(\ref{cut0}) was used). That is we may conclude that the
gravitational radiation from the cusp of the superconducting
chiral string depends on the current as follows:
\begin{equation}
\label{Eeps} E(\epsilon)=E(0)-G\mu^2\mathcal{B}\sqrt{\epsilon},
\end{equation}
where $\mathcal{B}$ is a numerical constant depending on the
string configuration ($G\mu^2$ was introduced in (\ref{Eeps}) for
dimension reasons). In the following we make the detailed
calculation of the radiation from the cusp that confirms of our
estimation (\ref{Eeps}) and gives the exact value of
$\mathcal{B}$ for any given configuration of the string.

The gravitational radiation from any relativistic source is given
\cite{Weinberg}
\begin{eqnarray}
\label{EW} \frac{d E}{d\Omega}\!\!&=&\!\!
2G\int_0^{\infty}\!\!\!\omega^2\,
d\omega\, P_{ij}P_{lm}\times\\
&\times&\!\!\left[T_{il}^*(\omega,\mathbf{k})T_{jm}(\omega,\mathbf{k})-
\frac{1}{2}T_{ij}^*(\omega,\mathbf{k})T_{lm}(\omega,\mathbf{k})
\right],\nonumber
\end{eqnarray}
where $P_{ij}=\delta_{ij}-n_i n_j$ is the projection operator on
the plane orthogonal to $\mathbf{n}$ and
$T_{jm}(\omega,\mathbf{k})$ is given by (\ref{TF}). The
expression (\ref{EW}) can be greatly simplified if we rewrite it
in 'co-rotating' basis $(\mathbf{n},\mathbf{v},\mathbf{w})$ where
$\mathbf{v}$ and $\mathbf{w}$ are the unit vectors, perpendicular
each other and to the vector $\mathbf{n}$ \cite{Durrer}. In a new
basis the radiated energy is
\begin{eqnarray}
  \label{ED}
  \frac{d E}{d\Omega}\!\!&=&\!\! 2 G\!
  \int_0^{\infty}\!\!\!\omega^2\,
d\omega\times \\
&\times&\!\!\left[{\tau}^{*}_{pq}(\omega,\mathbf{k})
{\tau}_{pq}(\omega,\mathbf{k})-
\frac{1}{2}{\tau}^{*}_{qq}(\omega,\mathbf{k}){\tau}_{pp}(\omega,\mathbf{k})
\right],\nonumber
\end{eqnarray}
where  $\tau_{pq}(\omega,\mathbf{k})$ are the Fourier-transforms
of an energy-momentum tensor in the co-rotating basis
$(\mathbf{n},\mathbf{v},\mathbf{w})$. Only indexes $p,q$ with
values $2$ and $3$ appear in the equation (\ref{ED}). The
Fourier-transforms $\tau_{pq}(\omega,\mathbf{k})$ can be
expressed in the following way
\begin{equation}
\label{tau}
\tau_{pq}(\omega,\mathbf{k})=-\frac{\mu}{4\pi}\hat{I}^{(p}\hat{Y}^{q)},
\end{equation}
where the 'modified' functions $\hat{I}_{p}$ and $\hat{Y}_{q}$
are determined as follows:
\begin{eqnarray}
  \label{IYmod}
  \hat{I}_{2}\!&\equiv&\! \hat{I}(\mathbf{n},\mathbf{v}),\quad
  \hat{I}_{3}\equiv \hat{I}(\mathbf{n},\mathbf{w}),\nonumber\\
  \hat{Y}_{2}\!&\equiv&\! \hat{Y}(\mathbf{n},\mathbf{v}), \quad
  \hat{Y}_{3}\equiv \hat{Y}(\mathbf{n},\mathbf{w}),
\end{eqnarray}
with
\begin{eqnarray}
  \label{IYmod1}
  \hat{I}(\mathbf{n},\mathbf{z})=
  \int_{-\infty}^{\infty} d\sigma_-
  (\mathbf{X}'_-\mathbf{z})e^{-\frac{i}{2}k X_-},
  \nonumber\\
  \hat{Y}(\mathbf{n},\mathbf{z})=
  \int_{-\infty}^{\infty} d\sigma_+ (\mathbf{X}'_+\mathbf{z})
  e^{-\frac{i}{2}k X_+}.
\end{eqnarray}
We extended the integration on $\sigma_{\pm}$ in (\ref{IYmod1})
from $-\infty$ to $\infty$ as the radiation mostly come from the
vicinity of the cusp. Inserting (\ref{tau}) in (\ref{ED}) we find:
\begin{equation}
\label{Efin} \frac{d E}{d\Omega}=\frac{G\mu^2}{8\pi^2}
\int_0^{\infty}\!\!\omega^2\,d\omega\,\mathcal{K},
\end{equation}
where
\begin{eqnarray}
\label{K} \mathcal{K}=\left|\hat{I}_2\hat{Y}_2\right|^2+
\left|\hat{I}_3\hat{Y}_3\right|^2
&+&\frac{1}{2}\left|\hat{I}_2\hat{Y}_3+\hat{I}_3\hat{Y}_2\right|^2\nonumber\\
&-&\frac{1}{2}\left|\hat{I}_2\hat{Y}_2+\hat{I}_3\hat{Y}_3\right|^2.
\end{eqnarray}
For the definiteness let us choose the following co-rotating
basis:
\begin{eqnarray}
\label{basis}
\mathbf{n}&=&(\cos\theta,\,-\sin\phi\sin\theta,\,
\cos\phi\sin\theta),\nonumber\\
\mathbf{v}&=&(-\sin\theta,\,-\sin\phi\cos\theta,\,
\cos\phi\cos\theta),\nonumber\\
\mathbf{w}&=&(0,\,\cos\phi,\,\sin\phi),
\end{eqnarray}
and $\mathbf{n}^{(c)}=(1,0,0)$. The difference between vectors
$\mathbf{n}$ and $\mathbf{n}^{(c)}$ is $\vec{\delta}$ so that
\begin{equation}
\label{delta} \mathbf{n}=\mathbf{n}^{(c)}+\vec{\delta}\simeq
\mathbf{n}^{(c)}+\theta\vec{\psi},
\end{equation}
where $\vec{\psi}=(-\theta/2,-\sin\phi,\cos\phi)$. In this basis
we have $\mathbf{w}\mathbf{n}^{(c)}\equiv 0$ therefore from
(\ref{IYmod1}) we find that ${\rm Re}\,\hat{I}_3= {\rm
Re}\,\hat{Y}_3\equiv 0$. Thus the expression for $\mathcal{K}$
(\ref{K}) can be rewritten in a following way:
\begin{eqnarray}
\label{K1} \mathcal{K}= \frac{1}{2}\left(|\hat{I}_2 |^2+
|\hat{I}_3|^2\right)
\left(|\hat{Y}_2|^2+|\hat{Y}_3|^2\right)\nonumber\\
+2{\rm Re}\,\hat{I}_2\,{\rm Im}\,\hat{I}_3\,{\rm Re}\,\hat{Y}_2\, {\rm
Im}\,\hat{Y}_3.
\end{eqnarray}
Let us denote further $\mathbf{X}''_{\pm}=\vec{\rho}_{\pm}/L$,
where $L$ is the invariant length of the string. Using the
relations (\ref{Taylor}), (\ref{xl}) and (\ref{delta}) after some
algebra one can find the expressions for phases
$\varphi_{\pm}=(\omega/2) n X_{\pm}$ (where the notation
$n^{(c)}=(1,\mathbf{n}^{(c)})$ was introduced) entering the
formulas for functions $\hat{I}_p$ and $\hat{Y}_q$ (\ref{IYmod1}):
\begin{eqnarray}
\label{kX} \varphi_-&=&\frac{\omega}{4}\left[\theta^2\sigma_-
+\frac{\theta}{L}(\vec{\psi}\vec{\rho}_-)\sigma_-^2
+\frac{1}{3}\frac{(\vec{\rho}_-)^2}{L^2}\sigma_-^3\right],\\
\varphi_+&=&-\frac{\omega}{4}\left[(\theta^2+2\epsilon)\sigma_+
-\frac{\theta}{L}(\vec{\psi}\vec{\rho}_+)\sigma_+^2
+\frac{1}{3\Delta}\frac{(\vec{\rho}_+)^2}{L^2}\sigma_+^3\right].\nonumber
\end{eqnarray}
Then we can evaluate the integrals in (\ref{IYmod1}) using the
formulas (in the analogy with analysis carried out in
\cite{Blanco}):
\begin{eqnarray}
\label{Bessel} \int_0^{\infty}\!d x\,
\cos\left[\frac{3}{2}\xi\left(x+\frac{x^3}{3}\right)\right]=
\frac{1}{\sqrt{3}}K_{1/3}(\xi),\nonumber\\
\int_0^{\infty}\! d x\, x\,
\sin\left[\frac{3}{2}\xi\left(x+\frac{x^3}{3}\right)\right]=
\frac{1}{\sqrt{3}}K_{2/3}(\xi),
\end{eqnarray}
where $K_{1/3}(\xi)$ and $K_{2/3}(\xi)$ are the modified Bessel
functions of order $1/3$ and $2/3$ correspondingly. Denoting
\begin{eqnarray}
\xi_-\!\!&=&\!\!\frac{1}{6}\omega\frac{L}{\rho_-}\theta^3\mathcal{Y}_-^3,
\quad
\chi_-=\frac{L}{\rho_-}\theta\mathcal{Y}_-,\nonumber \\
\xi_+\!\!&=&\!\!\frac{1}{6}\omega\frac{L}{\rho_+}
\left(\theta^2\mathcal{Y}_+^2+2\epsilon\right)^{3/2},\quad
\chi_+\!\!=\!\!\frac{L}{\rho_+}
\left(\theta^2\mathcal{Y}_+^2+2\epsilon\right)^{1/2},\nonumber
\end{eqnarray}
where
\begin{eqnarray}
\mathcal{Y}_{\pm}=
\frac{\sqrt{{\rho}_{\pm}^2-(\vec{\rho}_{\pm}\vec{\psi})^2}}{\rho_{\pm}},\nonumber
\end{eqnarray}
after some calculations one can obtain from (\ref{IYmod1}):
\begin{eqnarray}
\label{IYfin} \hat{I}_2&=&
\frac{2}{\sqrt{3}}\frac{L}{\rho_-}\theta^2\mathcal{Y}_-K_{1/3}(\xi_-)\nonumber\\
&-&i\frac{2}{\sqrt{3}}\frac{L}{\rho_-}
\frac{(\vec{\rho}_-\mathbf{v})}{\rho_-}\theta^2\mathcal{Y}_-^2K_{2/3}(\xi_-),\nonumber\\
\hat{I}_3&=&-i\frac{2}{\sqrt{3}}\frac{L}{\rho_-}
\frac{(\vec{\rho}_-\mathbf{w})}{\rho_-}\theta^2\mathcal{Y}_-^2K_{2/3}(\xi_-),\nonumber\\
\hat{Y}_2&=&
-\frac{2}{\sqrt{3}}\frac{L}{\rho_+}\theta\sqrt{\theta^2\mathcal{Y}_+^2+\epsilon}
K_{1/3}(\xi_+)\\
&-&i\frac{2}{\sqrt{3}}\frac{L}{\rho_+}
\frac{(\vec{\rho}_+\mathbf{v})}{\rho_+}
\left(\theta^2\mathcal{Y}_+^2+\epsilon\right)K_{2/3}(\xi_+),\nonumber\\
\hat{Y}_3&=&-i\frac{2}{\sqrt{3}}\frac{L}{\rho_+}
\frac{(\vec{\rho}_+\mathbf{w})}{\rho_+}
\left(\theta^2\mathcal{Y}_+^2+\epsilon\right)K_{2/3}(\xi_+)\nonumber.
\end{eqnarray}
The expressions for $\hat{I}_p$ and $\hat{Y}_q$ in the case of the
ordinary string can be found from (\ref{IYfin}) by setting
$\epsilon=0$. All we have to do is to insert the result
(\ref{IYfin}) in (\ref{K1}) and calculate the radiated power
using (\ref{Efin}). The expressions (\ref{IYfin}) was found under
the assumption that $\theta\ll 1$. In the case of the
electromagnetic radiation Blanco-Pillado  and Olum \cite{Blanco}
extended the obtained results for small $\theta$ to any values of
$\theta$, because the electromagnetic radiation is highly beamed
and integral for radiated power over $\theta$ converges vary
fast. This procedure can not be applied for the gravitational
radiation because the radiated gravitational power (\ref{estE})
does not converges for large $\theta$. Nevertheless we can find
the difference between the radiated powers of ordinary and chiral
cosmic string cusps in the case $\epsilon\ll 1$ using the fact
that the difference of powers arises for the $\theta\alt\epsilon$
where our calculations are valid. In other words in our analysis
we are able to calculate the value $\mathcal{B}$ in formula
(\ref{Eeps}) (which will be verified below) but we can not find
the value $E(0)$, i.e. the radiated power for the ordinary string
cusp.

\begin{figure}[t]
\includegraphics[width=0.48\textwidth]{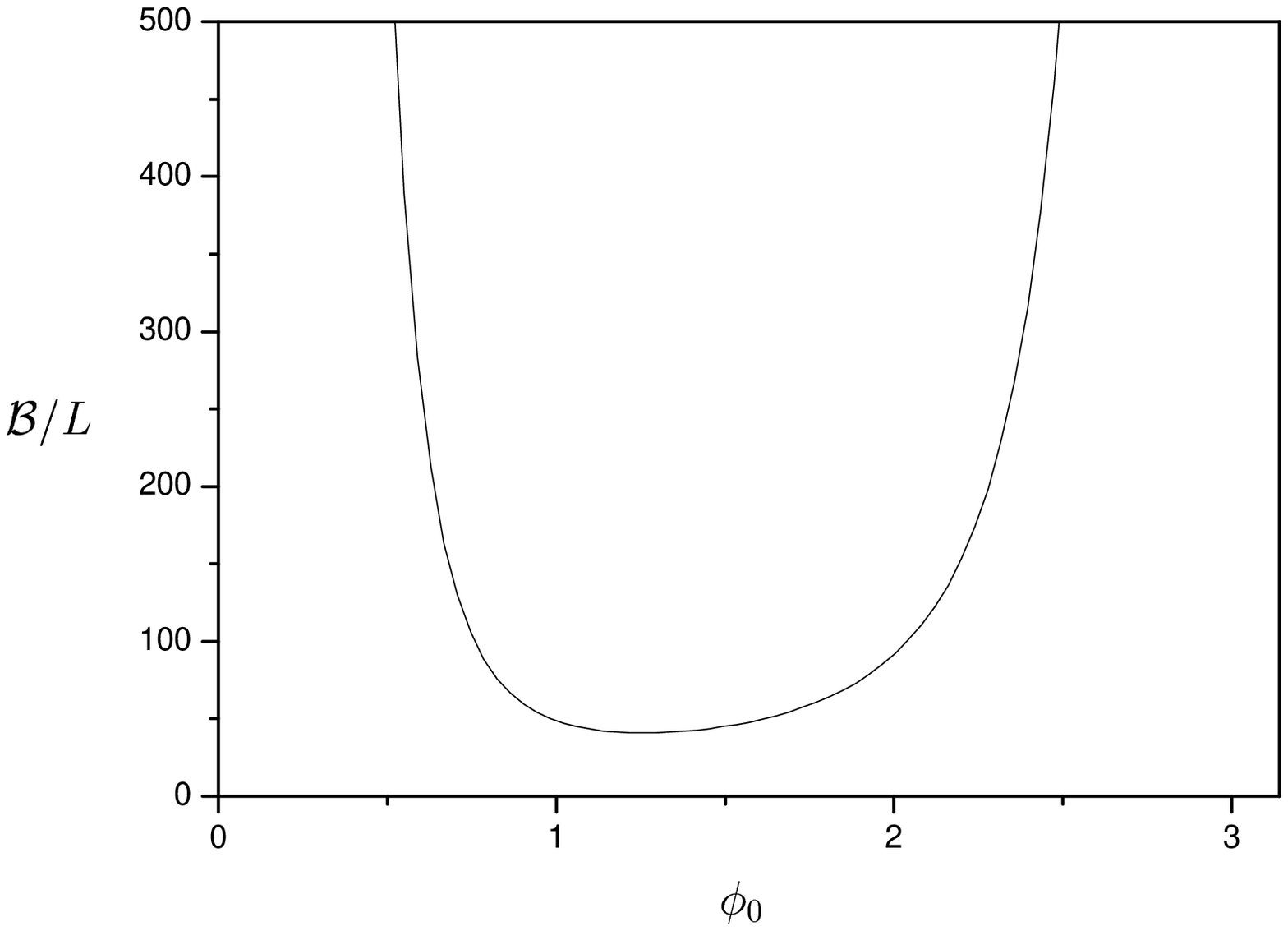}
\caption{\label{fig1} The coefficient $\mathcal{B}$ in (\ref{B}) as a
function of an angle $\phi_0$ which parameterizes the geometry of the
string loop (\ref{loops}).}
\end{figure}

Without the loss of the generality we can choose
$\vec{\rho}_-=\rho_-(0,0,1)$ and
$\vec{\rho}_+=\rho_+(0,-\sin\phi_0,\cos\phi_0)$ where $\phi_0$ is
the angle between $\vec{\rho}_-$ and $\vec{\rho}_+$. Then we have
$\mathcal{Y}_-=|\sin\phi|$ and
$\mathcal{Y}_+=|\sin(\phi-\phi_0)|$. Substituting (\ref{IY}) into
(\ref{K1}), then obtained expression into (\ref{Efin}) and making
the successive change of variables:
\begin{equation}
\omega=\frac{6}{\mathcal{Y}_-^3\theta^3}\frac{\rho_-}{L}z, \quad
\theta=\theta'\sqrt{\epsilon},
\end{equation}
we finally find:
\begin{widetext}
\begin{equation}
\label{Fin} E(\epsilon)=E(0)
-\sqrt{\epsilon}\,\frac{48}{\pi^2}\frac{\rho_-^{1/3}}{\rho_+^{4/3}}G\mu^2
L \int_0^{2\pi}\frac{d\phi}{\mathcal{Y}_-^5}
\int_0^{\infty}d\theta'\left[\mathcal{G}(a)-\mathcal{G}(a_0)\right],
\end{equation}
where
\begin{eqnarray}
\label{G} \mathcal{G}(a)\!\! &=& \!\!\int_0^{\infty}\!\!d z\, z^2
a^{2/3}\left\{\frac{1}{2}\left[K_{1/3}^2(z)+\mathcal{Y}_-^2K_{2/3}^2(z)\right]
\left[K_{1/3}^2(a z)+\mathcal{Y}_-^2
a^{2/3}\left(\frac{\rho_+}{\rho_-}\right)^{2/3}
\mathcal{Y}_-^2K_{2/3}^2(a z)\right] \right. \nonumber\\
&&+\left.
2\frac{\mathcal{Y}_-^2(\vec{\rho}_-\vec{\psi})(\vec{\rho}_+\vec{\psi})
a^{1/3}}{\rho_-\rho_+}\left(\frac{\rho_+}{\rho_-}\right)^{1/3}
K_{1/3}(z)K_{2/3}(z)K_{1/3}(a z)K_{2/3}(a z), \right\}
\end{eqnarray}
\end{widetext}
with
\begin{eqnarray}
\label{a} a &=& \frac{\rho_-}{\rho_+}
\left[\frac{\theta^2\sin^2(\phi-\phi_0)+2}
{\theta^2\sin^2(\phi)}\right]^{3/2}, \nonumber \\
a_0 &=&
\frac{\rho_-}{\rho_+}\left|\frac{\sin(\phi-\phi_0)}{\sin(\phi)}\right|^{3}.
\end{eqnarray}
We see that obtained result confirm our estimation (\ref{Eeps}) with:
\begin{eqnarray}
\label{B}
\mathcal{B}=\frac{48}{\pi^2}\frac{\rho_-^{1/3}}{\rho_+^{4/3}}L
\int_0^{2\pi}\frac{d\phi}{\mathcal{Y}_-^5}
\int_0^{\infty}d\theta'\left[\mathcal{G}(a)-\mathcal{G}(a_0)\right].
\end{eqnarray}
The dependence of radiated energy (\ref{Eeps}) on $\epsilon$ can
be expressed in terms of the physical current $j$, using
(\ref{current}):
\begin{equation}
\label{Ej} E(j)=E(0)-\frac{G\mu^2\mathcal{B}}{\sqrt{2\mu}}j
\end{equation}

On the Fig. {\ref{fig1}} the calculated values of $\mathcal{B}/L$
as a function of $\phi_0$ is shown for the following particular
loop:
\begin{eqnarray}
\label{loops}
\mathbf{a}\!\!\!&=&\!\!\!\frac{L}{2\pi}\left(-\sin{\frac{2\pi\sigma_-}{L}}\,,0\,,\,
-\cos{\frac{2\pi\sigma_-}{L}}\right),\nonumber\\
\mathbf{b}\!\!\!&=&\!\!\!\frac{L}{2\pi}\times\\
\!\!\!&\times&\!\!\!\left(\sin{\frac{2\pi\sigma_+}{L}},
\,\sin\phi_0\cos{\frac{2\pi\sigma_+}{L}}\,,
-\cos\phi_0\cos{\frac{2\pi\sigma_+}{L}}\right).\nonumber
\end{eqnarray}
Note that the function $\mathcal{B}$ is not reflection symmetric with
respect to the point $\phi_0=\pi/2$. This is because of a presence of the
term proportional to $\cos\phi\,\cos(\phi-\phi_0)$ in (\ref{G}).

\section{Conclusion}

The gravitational radiation from the chiral string cusp is considered
analytically in the limit of the small current, $\epsilon=1-\Delta\ll 1$,
where $\Delta$ is from (\ref{ab}). We showed that if the back-reaction of
charge carriers is taken into account then the chiral string cusp (and
consequently the whole string) does not radiate the gravitational energy
at frequencies $\omega\agt\omega_{\rm cut}$. The cutoff frequency
$\omega_{\rm cut}$ depends on the current and is defined by (\ref{cut0}).
If the current on the string is sufficiently large then this cutoff may
lead to an effective lowering of the amplitude of GWBs. This would affect
the possible detection of GWBs coming from the whole universe by the
interferometric gravitational detectors LIGO, VIGRO and especially LISA.
The critical value of the chiral current (when the smoothing of the cusp
becomes to influence the detectability) depends according to (\ref{ecut})
on the observation frequency $\omega$, assumed rate of GWBs $\dot N$ and
the mass parameter of the string $\mu$ . For example, for GUT string with
$G\mu\sim 10^{-6}$ the critical current $j=\sqrt{2
\mu}\sqrt{\epsilon_{\rm cut}}\sim 10^{11}$ GeV, which corresponds to
$\epsilon_{\rm cut}\sim 10^{-9}$.

The analytical dependence of the total gravitational energy radiation from
the cusp as a function of the small current was calculated. Namely it was
found that the radiation decreases with the current on the string as
follows: $E(\epsilon)=E(0)-G\mu^2 \mathcal{B}\sqrt{\epsilon}$. For the
one-parametric family of string configurations (\ref{loops}) we found the
numerical values of $\mathcal{B}\sim 50\div 400$.

\begin{acknowledgments}
This work was supported in part by Russian Foundation for basic Research
grants 02-02-16762-a and by the INTAS through grant 99-1065.
\end{acknowledgments}

\end{document}